\documentclass[aps, pre, twocolumn]{revtex4-1}
\usepackage[utf8x]{inputenc}
\usepackage{epsfig}
\usepackage{amsmath,latexsym}
\usepackage{color}
\usepackage{xcolor, soul}
\usepackage[toc,page]{appendix}
\usepackage{graphicx}
\usepackage{dcolumn}
\usepackage{bm}
\usepackage[normalem]{ulem} 

\newcommand{\rfrac}[2]{{}^{#1}\!/_{#2}}

\begin{document}
\title{Experiments and characterization of low-frequency oscillations in a granular column}
\author{Loreto Oyarte G\'alvez$^{1}$} \email{l.a.oyartegalvez@utwente.nl}
\author{Nicol\'as Rivas$^{2}$}
\author{Devaraj van der Meer$^{1}$}
\affiliation{$^{1}$Physics of Fluids, Universiteit Twente, Post Office Box 217, 7500AE Enschede, The Netherlands}
\affiliation{$^{2}$Forschungszentrum J{\"u}lich GmbH, Helmholtz-Institut Erlangen-N{\"u}rnberg f{\"u}r Erneuerbare Energien (IEK-11), Dynamik komplexer Fluide und Grenzfl{\"a}chen, F{\"u}rther Straße 248, 90429 N{\"u}rnberg, Germany}

\date{\today}
 
\begin{abstract} 
The behaviour of a vertically vibrated granular bed is reminiscent of a liquid in that it exhibits many phenomena such as convection and Faraday-like surface waves. However, when the lateral dimensions of the bed are confined such that a quasi-one-dimensional geometry is formed, the only phenomena that remain are bouncing bed and the granular Leidenfrost effect. This permits the observation of the granular Leidenfrost state for a wide range of energy injection parameters, and more specifically allows for a thorough characterisation of the low-frequency oscillation (LFO) that is present in this state. In both experiments and particle simulations we determine the LFO frequency from the power spectral density of the centre of mass signal of the grains, varying the amplitude and frequency of the driving, the particle diameter and the number of layers in the system. We thus find that (i) the LFO frequency is inversely proportional to the fast inertial time scale and (ii) decorrelates with a typical decay time proportional to the slow dissipative time scale in the system. The latter is consistent with the view that the LFO is driven by the inherent noise that is present in the granular Leidenfrost state with a low number of particles. \\ 
\end{abstract}
  
\maketitle
\section{Introduction}\label{s:Intro}

Granular materials are observed on a daily basis; they are present in many natural phenomena such as avalanches, land-slides or the formation of dunes, and they have a wide range of applications \cite{jaeger_physics_2008}. The exploration of the similarities between the behaviour of granular materials and that of ordinary fluids has motivated much appealing research. An important example is a vertically shaken granular bed, which exhibits fluid-like behaviour \cite{eshuis_phase_2007}  {that in turn depends}, in part, on the injected energy, the number of particle layers and the system geometry. As the shaking energy is increased, the system transits from (i) a bed of grains bouncing with the base; to (ii) bursts; (iii) undulations \cite{douady_subharmonic_1989, clement_granular_1998, sano_dilatancy_2005, melo_transition_1994,moon_phase_2001} (which are analogous to Faraday waves in regular liquids \cite{faraday_peculiar_1831, cross_pattern_1993}); (iv) density inversion~\cite{eshuis_granular_2005,lim_granular_2010,roeller_dynamics_2012}, where a dense layer of grains floats on top of a gaseous layer, a state referred to as the Leidenfrost state due to its similarity to a liquid droplet floating on its own vapour above a hot plate~\cite{leidenfrost_aquae_1756};  (v) buoyancy driven convection rolls~\cite{chandrasekhar_hydrodynamic_1961, paolotti_thermal_2004} and, for very high shaking energy; to (vi) a granular gas (a dilute granular system with particles moving randomly throughout the container). 

For a sufficiently large number of particle layers the granular bed transits from the Leidenfrost state to the convection state as the energy input is increased. Recently, an oscillation was observed in the motion of the dense part of the Leidenfrost state~\cite{rivas_low-frequency_2013, wakou_fluctuation-dissipation_2012}. The frequency of these oscillations is typically much lower than the frequency of the injected energy, and becomes dominant in the dynamics of the system when the energy input increases. Recently, these so-called low-frequency oscillations (LFO) were indirectly observed in a three-dimensional vibrofluidized granular bed, by tracing the movement of a single particle through the bulk using positron emission particle tracking~\cite{windows-yule_low-frequency_2014}, showing a good agreement with numerical simulations, where the effect was first observed~\cite{rivas_low-frequency_2013}.

\begin{figure}[h!]
  \begin{center}
   \includegraphics[scale=1]{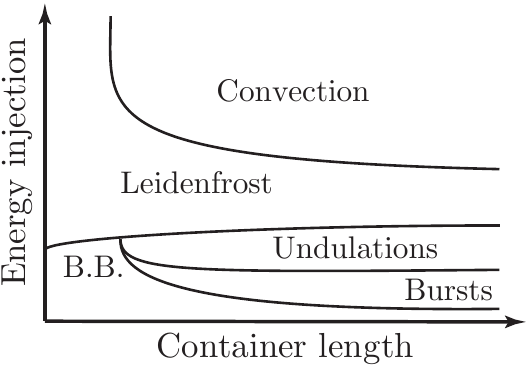}
  \end{center}
  \vspace{-0.2cm}
   \caption{Schematic phase diagram of a vibrated quasi-two-dimensional granular bed as a function of the energy injection and the container's length. Here, B.B. indicates the bouncing bed state and the granular gas state is reached for very high energy injection which lies outside the range of the phase diagram.}
  \label{fig:PhaseDiagram}
\end{figure}

How the above fluid-like behaviours in a granular bed depend on the container's geometry can be represented in a phase diagram, as shown in figure \ref{fig:PhaseDiagram}~\cite{rivas_low-frequency_2013}. Here the occurrence of the phenomena is schematically indicated as a function of the container width and the energy injection. When the container is large enough (larger than approximately 20 particle diameters) and filled with a sufficiently large amount of particles, the system can exhibit any of the above mentioned behaviours depending on the energy injection parameters. But for narrower containers many of these phenomena are suppressed by the geometry of the container, since bursts, undulations and convection rolls are extended in the horizontal direction. These patterns are frustrated by the presence of the side walls, which can be attributed to e.g., effective viscosity and lateral heat conduction~\cite{bromberg_development_2003}. Therefore, in a quasi-one-dimensional container only the bouncing bed and Leidenfrost states are present. 

The possibility to keep the system in the Leidenfrost regime  {by using a narrow container} and thus observe LFO's for a wide range of shaking strengths has motivated this work. We experimentally study the behaviour of the LFO's and their role in the dynamics of the system. The results are compared with simulations showing a very good agreement between both. Moreover, we analyse the time scales present in the system and relate them to the LFO frequency, finding that when using suitable dimensionless quantities the LFO frequencies collapse onto a single curve independent of the number of layers and particle diameter. Finally, we analyse the strength of the LFO by describing the system by means of a Langevin equation of a noise-driven harmonic oscillator. We find that the measured LFO's strength and the strength scaling derived from the Langevin equation are consistent, and we show that the energy equipartition is increasingly violated with the increment of the shaking strength.    

This article is organised as follows. In section \ref{s:expSetup}, the experimental setup and the simulations are detailed. In section \ref{s:Analysis}, we analyse the experimental results and compare them to the simulations; in addition the different time scales presented in the system are explained and analysed. Finally, in section \ref{s:Conclusions} a summary of this study is presented. 

\section{System Description}\label{s:expSetup}

\subsection{Experimental Setup}
\begin{figure}[h!]
   \begin{center}
   \includegraphics[scale=1]{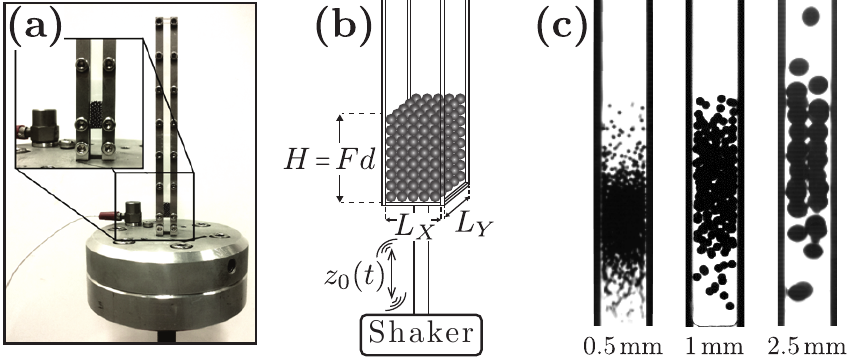} 
   \end{center}
   \vspace{-0.2cm}
   \caption{(a) A full picture and a zoomed-in view of the experimental setup: The container is mounted on top of the mechanical shaker, with the bottom located at approximately $1$ cm above the socket. Note that the granular material above it is at rest. (b) A schematic of the container, where the dimensions of the container and the filling height $H$, related to the filling factor $F$ and particle diameter $d$, are indicated. (c) Typical images of the partially filled container under shaking, for the three different particle diameters, $d$=0.5, 1.0, 2.5[mm].}
   \label{fig:ExperimentalSetupZoom}
\end{figure}

The experimental setup consists of a quasi-one dimensional transparent acrylic container, with base dimensions ($L_X,\,L_Y$)  much smaller than the height (height$\times L_X\times L_Y$ = 150$\times$5$\times$5~[mm$^3$]), as shown in figure \ref{fig:ExperimentalSetupZoom} (a). The container is partially filled with mono-dispersed stainless steel beads of three different diameters, $d = 0.5$, $1.0$ and $2.5$ [mm], i.e. the container width $L=L_X=L_Y$ corresponds to $L=10d,$ $5d$ and $2d$ respectively. The system is mounted on a sinusoidally vibrating electromechanical shaker with tuneable frequency $f_0$ and amplitude $a_0$, i.e., the vertical position of the bottom is given by
\begin{equation}
  z_0(t)=a_0\sin(2\pi f_0 t).
\end{equation}

Front view images are obtained using a high-speed camera capturing 500 frames per second during 2 minutes; examples for every particle size are shown in figure \ref{fig:ExperimentalSetupZoom} (c). For every set of parameters $5$ to $10$ acquisitions were performed, in order to reduce the statistical error. 

Two dimensionless numbers have been previously found to be relevant in the Leidenfrost state \cite{eshuis_granular_2005}, which we choose as control parameters: i) the number of monolayers at rest $F=Nd^2/L^2$, with $N$ the total number of particles; and ii) the dimensionless shaking strength $S=(2\pi a_0f_0)^2/g\ell$, where $\ell$  corresponds to the typical displacement of the particles and $g$ is the gravitational acceleration. Both dimensionless numbers, $F$ and $S$, are varied by changing the parameters $a_0$, $f_0$,  {$N$} and $d$. 

\begin{figure}[h!]
  \begin{center}
   \includegraphics[scale=1]{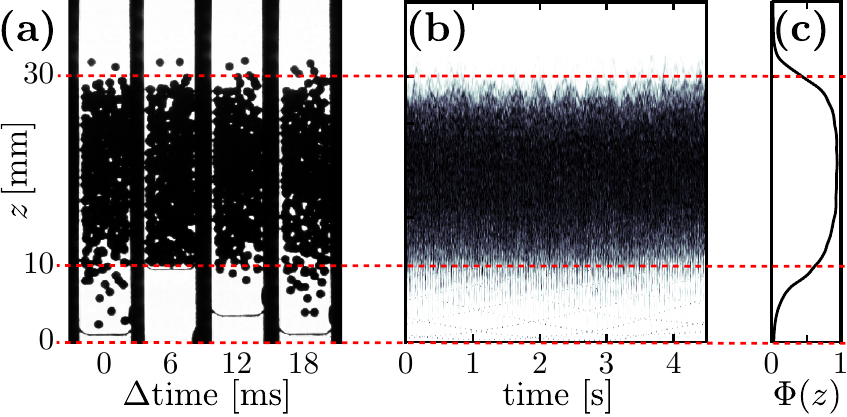}
  \end{center}
  \vspace{-0.2cm}
  \caption{(a) A sequence of images of the system in different phases of the driving showing a dense cluster floating over a dilute, gaseous granular layer. This sequence corresponds to one period of oscillation of the shaker. (b) Average of the black and white pixels on every horizontal level for each frame as a function of time, referred to as density profile. (c) Time-averaged density profile of the system.}
  \label{fig:rhovsTime}
\end{figure}

\begin{figure}[h!]
  \begin{center}
   \includegraphics[scale=1]{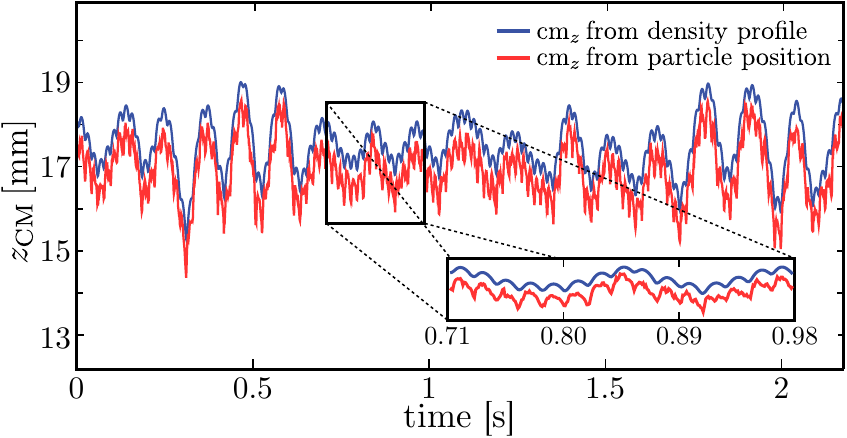}
  \end{center}
  \vspace{-0.2cm}
  \caption{A comparison of the vertical position of the centre of mass $z_\text{CM}(t)$ computed directly from the position of the particles (blue line) and using the experimental method, i.e., from the density profile (red line); data obtained using numerical simulations. The frequency that dominates the main plot corresponds to that of the LFO, whereas in the region that is magnified in the inset the driving frequency $f_0$ can also be appreciated.}
  \label{fig:CMComp}
\end{figure}

In figure \ref{fig:rhovsTime} (a), a sequence of images shows the system in the Leidenfrost regime, where a dense volume of grains is seen to float over a much less dense gaseous layer. From the images it is not possible to obtain the position of all the particles as it is not feasible to distinguish individual particles for such high densities. Therefore, in order to obtain the vertical coordinate of the centre of mass we compute the horizontal intensity average of every image, obtaining a vertical density profile  at each point in time, as shown in figure \ref{fig:rhovsTime} (b). We then define the vertical positions of the centre of mass $z_\text{CM}(t)$ of the granular bed as that of the centre of mass of the density profile at each individual point in time (cf. figure \ref{fig:rhovsTime} (b) and (c)). To verify this procedure we use data from numerical simulations (detailed further below) to compute the $z_\text{CM}(t)$ using two approaches: (i) from the exact position of the particles, and (ii) from the density profile as done in experiments. As shown in figure \ref{fig:CMComp}, there is a quite good agreement in the computed $z_\text{CM}(t)$ using both approaches. However, the density profile approach consistently over-predicts $z_\text{CM}(t)$ since it does not differentiate between a single particle in the gaseous regime and a series of particles stacked in the Y-direction in the dense region. Nonetheless, the frequency of oscillation of the $z_\text{CM}(t)$ is accurately captured.

\subsection{Simulations}

Simulations are performed using event-driven (ED) molecular dynamics \cite{lubachevsky_how_1991}. Particles are considered as hard spheres, which implies binary collisions, no overlap and no long-range forces between them. Collisions are modelled by normal and tangential velocity-dependent restitution coefficients, following the expression in Ref. \cite{mcnamara_simulations_2005}. This is the same simulation code used in the original study of LFO \cite{rivas_low-frequency_2013}, where a more detailed description of the algorithm can be found. Material properties were chosen such that, at a typical particle velocity $\bar{v} = 0.3$[m/s], the relevant coefficient of restitution is $\bar{r} = 0.93$ for both particle-particle and particle-wall collisions. Velocity-dependent $\bar{r}$ ensures that dissipation is not overestimated at high particle densities, as can occur when using constant coefficients \cite{poschel_long-time_2003}. Static and dynamic friction coefficients ($\mu_s$ and $\mu_d$, respectively) are also considered, and held constant at $\mu_s = \mu_d = 0.08$ also for both types of collisions. In general, variation of these parameters influences the values of the measured quantities (as will be discussed later), but the qualitative aspects remain the same, even if periodic boundary conditions are used and friction is set to zero. These particular values of $\mu$ and $\varepsilon$ where taken as fitting parameters to obtain a good agreement with the experimental data, and are in the range of measured values for milimetric stainless steel spheres.

\subsection{Shaking strength} 

 The definition of the typical particle displacement $\ell$, and hence the dimensionless shaking strength $S$, is not straightforward, since it sensitively depends on the system state. When the granular bed is only slightly fluidised the typical displacement is highly correlated with the shaking amplitude, therefore $\ell=a_0$ is a good approximation, and the shaking strength becomes the dimensionless acceleration 
\begin{equation}
  S_{a_{_0}}=\Gamma=\frac{a_0(2\pi f_0)^2}{g} .
\end{equation}

On the other hand, when the granular bed is strongly fluidised this length scale is decoupled from the shaking amplitude $a_0$ and therefore an intrinsic parameter, namely the particle diameter $\ell=d$, becomes a more sensible choice~\cite{ eshuis_phase_2007,eshuis_granular_2005,pak_surface_1993}. With this choice, the dimensionless shaking strength becomes
\begin{equation}
  S_d=\frac{(2\pi a_0 f_0)^2}{d\,g} .
  \label{eq:Sd}
\end{equation}

For states with a high energy injection rate, such as the Leidenfrost regime, where it is clearly observed that particles near the bottom typically travel distances which are larger than the shaking amplitude,  $S_d$ appears as the proper dimensionless parameter to describe such a driven system (as long as the particle diameter $d$ is kept constant), as confirmed by experiments and a theoretical hydrodynamic model~\cite{eshuis_phase_2007}. In the following Section, we will however see that in our case, where $d$ is actually varied, we will need to reconsider the choice of the dimensionless control parameter.  

\section{Analysis of Results}\label{s:Analysis}

The granular Leidenfrost state is an example of spontaneous segregation or symmetry breaking where an initially homogeneous, monodisperse granular material separates into a dense and a dilute region. This is the result of interparticle collisions being dissipative and of stochastic nature~\cite{serero_hydrodynamics_2015,gunkelmann_stochastic_2014,gunkelmann_temperature_2013,Mujica_2012}. More specifically, in the granular Leidenfrost state a dense, liquid-like or even almost solid-like cluster is floating on top of a dilute, gaseous region. Such a system is analogous to a piston (the cluster) that encloses an amount of gas. Since the piston is free to move up and down, the enclosed gas responds as a spring: when the cluster moves down, it is compressed and tries to push the cluster upwards, thereby trying to restore the equilibrium situation. Clearly, when the cluster moves upwards, the gas expands, the force on the cluster decreases, and the equilibrium is again restored. Qualitatively this is the phenomenon that lies at the basis of the Leidenfrost oscillation (LFO). As mentioned before, and as was observed in previous works \cite{wakou_fluctuation-dissipation_2012,rivas_low-frequency_2013,windows-yule_low-frequency_2014}, by studying a quasi-one-dimensional geometry most of the phenomena in the rich phase diagram can be suppressed, leaving only the bouncing bed and Leidenfrost state to develop. This allows us to directly observe these collective oscillations of the particles in a relatively large parameter window. In this Section we will first study how the frequency $f_\text{LFO}$ of the Leidenfrost oscillation depends on the parameters of the system, where we predominantly vary the particle size $d$, the filling factor $F$ and the dimensionless shaking strength $S_d$. From an analysis of the relevant time scales in the system we subsequently determine a natural scaling for $f_\text{LFO}$ and show that with an appropriate choice of dimensionless parameters the data can be made to collapse onto a single curve. Subsequently, we study the coherence time and the amplitude of the oscillation and show that the mechanism that drives the LFO is connected to the stochastic character of the system.     

\subsection{LFO frequency} \label{ss:Analysis_LFO}
 
\begin{figure}[h!]
  \begin{center}
   \includegraphics[scale=1]{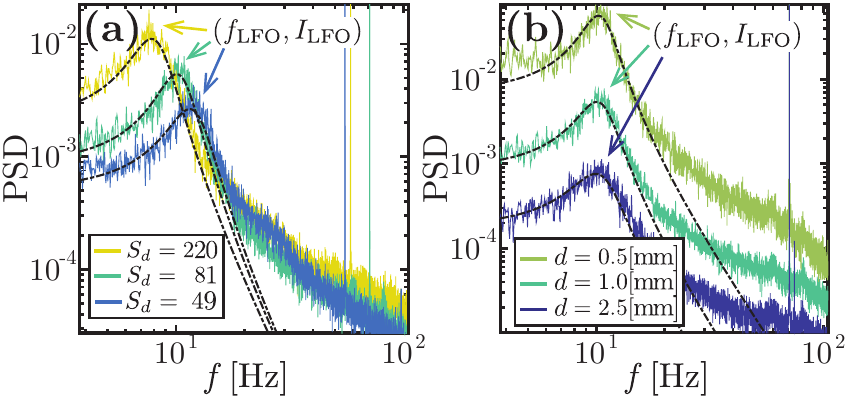}
  \end{center}
  \vspace{-0.2cm}
  \caption{The power spectral density (PSD) of the time evolution of the centre of mass $z_\text{CM}(t)$ for (a) different shaking strengths $S_d$ and (b) different particle diameters $d$, obtained from experimental data for $F=12$. The Leidenfrost oscillation (LFO) is clearly distinguished as a broad peak around the $10$[Hz] region and its frequency $f_\textrm{LFO}$ can be determined from the location of the peak's maximum. The black dashed lines represent theoretical results calculated from the Langevin model (Eq. (\ref{eq:PSD}))}
  \label{fig:PSD}
\end{figure}

 As can be seen in figure~\ref{fig:CMComp}, the time evolution of the centre of mass $z_\text{CM}(t)$ shows oscillations with two clearly distinct frequencies: a fast one, corresponding to the frequency of the shaker $f_0$, and a much  slower one, $f_\text{LFO}$, which corresponds to the low-frequency Leidenfrost oscillations. To determine this last frequency we compute the power spectral density (PSD) of $z_\text{CM}(t)$, shown in Figure~\ref{fig:PSD} for (a) different shaking strengths $S_d$  and (b) different particle diameters $d$. The frequency from the shaker $f_0$ and its harmonics are immediately recognised as well-defined, narrow peaks in the PSD. In addition,  there is an equally clear wide and shallow peak, which corresponds to the LFO. Surprisingly, the frequency $f_\textrm{LFO}$ decreases when the injected energy increases, in other words, the typical period of the collective oscillation of the system is larger when the injected energy increases. Furthermore, the  peak corresponding to the LFO becomes narrower and higher with increasing injected energy and with decreasing particle diameter. This, suggesting that this collective behaviour becomes more coherent in time as $S_d$ increases and $d$ decreases.  

\begin{figure}[h!]
  \begin{center}
   \includegraphics[scale=1]{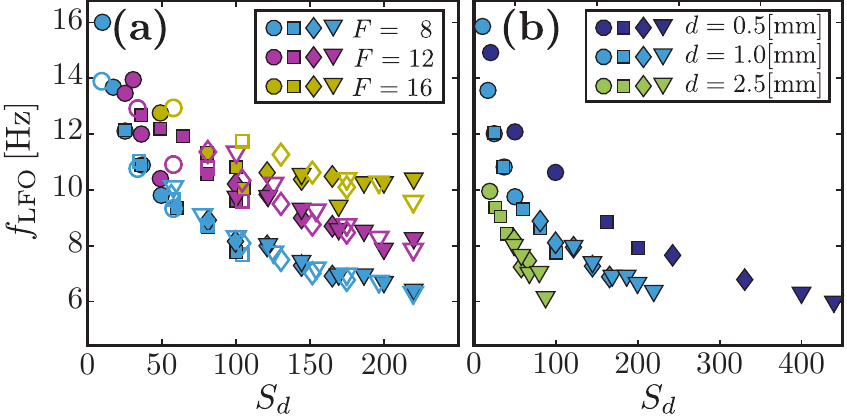}  
   \end{center}
   \vspace{-0.2cm}
  \caption{The LFO frequency $f_\textrm{LFO}$ is plotted versus the dimensionless shaking strength $S_d$ for (a) different numbers of layers $F$  and a constant particle diameter $d=1$[mm], and (b) different particle diameters and a constant number of layers $F=8$. The different symbols correspond to different amplitudes of the shaker, \Large{$\circ$}\small{$=1.0$[mm]}, $\Box=2.0$[mm], \Large{$\diamond$}\small{$=3.0$[mm]}, \small{$\bigtriangledown$}=4[mm]. The solid symbols represent experimental data and the open symbols numerical simulation results.}
  \label{fig:LFOvsS}
\end{figure}

The $f_\text{LFO}$ was determined from the PSD as a function of the driving strength $S_d$, for different numbers of layers $F$ and particle diameters $d$,  {the result of which is} shown in figure \ref{fig:LFOvsS} {(a) and (b) respectively.} The plot contains  {both experimental (closed symbols) and numerical (open symbols) data, where} the different symbols correspond to different values of the vibration amplitude $a_0$. The fact that the data collapse irrespective of the value of $a_0$ proves that $f_\text{LFO}$ indeed only depends on the combination $S_d$ and not on $a_0$ and $f_0$ separately. Our data is consistent with the scaling proposed by Rivas et al \cite{rivas_low-frequency_2013},  {namely }$f_\text{LFO}(S_d) \sim S_d^{-\alpha}$, with $\alpha \approx 0.3$.\\

 \subsection{A simple model for the LFO}
 
 A key question we want to address is: how does the frequency of the LFO scale with the parameters in the system? We find an answer by turning to a highly simplified mass-spring model of the Leidenfrost state (cf.~\cite{rivas_low-frequency_2013}, where a more elaborate model was introduced in the same spirit).

The dense high packing fraction region  at the top acts as a solid plug with a total mass $M_\text{plug}$, while the gaseous region below is equivalent to a spring, which applies a force to the plug that is proportional to its compression.  Once the spring constant $k$ is known, the frequency of oscillation of the spring is obtained from the force balance $M_\text{plug}\ddot{u}=-ku$, where $u$ is the vertical displacement of the plug with respect to its equilibrium position. This leads to the well-known  
relation $f_\textrm{LFO} = 2 \pi \sqrt{k/M_\textrm{plug}}$.  

To  estimate $k$ we use the fact that in steady state the pressure of the plug $P_\text{plug}$ must be equal to the pressure $P_\text{gas}$ in the gaseous region just below the plug
\begin{equation}
P_\text{gas}=P_\text{plug}=\frac{g M_\text{plug}}{L^2},
\end{equation}
\noindent where $L^2$ is the bottom area of the container. Using the (admittedly bold) assumption that the gaseous region is compressed in accordance with Boyle's law for ideal gases, we write $V_\text{gas} \Delta P_\text{gas} =  - P_\text{gas} \Delta V_\text{gas}$, where the equilibrium gas volume $V_\text{gas}$ is the typical height $\lambda$ times the  {bottom} area of the container $L^2$, the differential volume can be written as $\Delta V_\text{gas} = u L^2$, and the differential gas pressure $\Delta P_\text{gas}$ is written as the differential gas force per area $\Delta F_\text{gas}/L^2$.  This leads to 
\begin{equation}
  \Delta F_\text{gas} = - \left(\frac{g M_\text{plug}} {\lambda}\right) u,
\end{equation}
\noindent which is recognised as the force term in the equation for an harmonic oscillator with $k = g M_\textrm{plug} / \lambda$. 

The size of the gaseous region is proportional to the typical kinetic energy gain of a particle when colliding with the bottom, and gravity, i.e., $\lambda \sim T_0/g$, where $T_0$ will (at least for a constant filling factor $F$) be set by the temperature of the vibrating bottom $T_0 \sim a_0^2 f_0^2$. Inserting this approximation in the above expressions for $k$ and $f_\textrm{LFO}$ we find that $M_\text{plug}$ cancels out and
\begin{equation}
	f_\text{LFO}\sim\frac{g}{T_0^{\rfrac{1}{2}}}\sim \frac{g}{a_0f_0} \,\,\,\Rightarrow\,\,\,\widetilde{f}_\text{LFO} \equiv \frac{f_\text{LFO}a_0f_0}{g}
\end{equation}
\noindent providing us with a natural frequency scale for the LFO which directly leads to the above definition of the dimensionless LFO frequency $\widetilde{f}_\text{LFO}$.

The control parameter $S_d$ however also requires some thought. In fact, the choice of the particle diameter $d$ as a typical intrinsic length scale in equation~(\ref{eq:Sd}) had so far been motivated by the fact that it was a convenient length scale which was kept fixed in the work in which it was introduced. Since we actually vary $d$ we are in need of a control parameter that better corresponds to the physics of the system we are studying. Let us therefore make a few assumptions about the granular Leidenfrost state. First let us assume that the dissipation in the gaseous region is negligible in comparison with that in the dense region above it. This can be motivated from the fact that particles in the latter region are very close and therefore collisions very frequent. In a steady state this means that the particles in the gaseous region are just transporting energy from the bottom to the plug, where it is subsequently dissipated. This is reminiscent of famous a work from the Kadanoff group~\cite{du_1995,eshuis_2009}  where a single particle transports energy from a hot wall to a granular cluster. Now, in our case suppose that a single particle picks up a kinetic energy $\sim T_0 \sim a_0^2f_0^2$ at the bottom. This energy needs to be dissipated in the plug, and since in height there are typically $F$ particles to do so, the dissipated energy should scale as $F^2T_0 \sim (Fa_0f_0)^2$. It is therefore plausible that for Leidenfrost states with a similar gaseous region (and therefore similar $f_\textrm{LFO}$) we have the same value of  $Fa_0f_0$. Non-dimensionalizing the latter parameter with the velocity scale $\sqrt{gL}$ we obtain the following control parameter~\footnote{Note that this velocity scale has been chosen for convenience since the bottom dimensions $L$ are not varied in this experiment. One might equally well have chosen any other fixed length scale, such as a reference particle diameter.}
\begin{equation}
 B =\frac{a_0f_0F}{\sqrt{gL}}\,.
 \label{eq:defB}
 \end{equation}
     
\begin{figure}[h!]
  \begin{center}
  \includegraphics[scale=1]{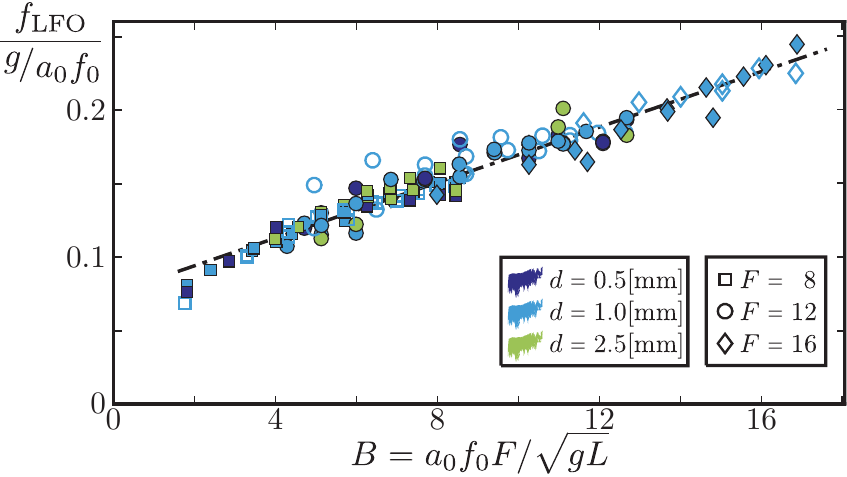}
  \end{center}
  \vspace{-0.2cm}
  \caption{The dimensionless frequency of the low-frequency oscillations $\widetilde{f}_\text{LFO}$ as a function of the shaking parameter $B$  collapses all simulational data (as shown in figure~\ref{fig:LFOvsS}) onto a single curve. Different symbols correspond to different numbers of layers $F$ and different colors to different particle diameters $d$.  As in figure~\ref{fig:LFOvsS}, the solid symbols represent experimental data and the open symbols numerical simulation results. The dashed-dotted line serves as a guide to the eye.}
\label{fig:LFOvsSHL} 
\end{figure}

In figure~\ref{fig:LFOvsSHL} the data of figure~\ref{fig:LFOvsS} is plotted using the new dimensionless parameters $\widetilde{f}_\text{LFO}$ and $B$. We see that all realised experiments and numerical simulations, collapse onto a single curve.\\
 
 \subsection{Time autocorrelation}
 
\begin{figure}[h!]
  \begin{center}
    \includegraphics[scale=1]{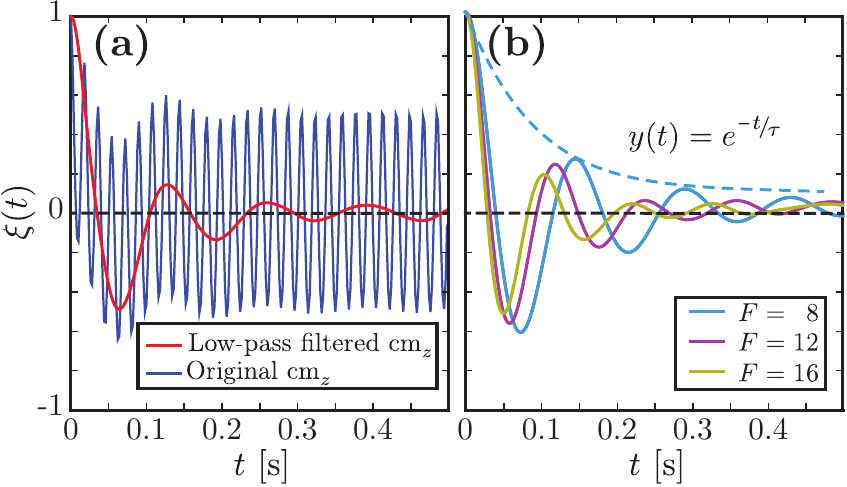}
  \end{center}
  \vspace{-0.2cm}
  \caption{(a) Autocorrelation function of the original centre of mass signal $z_\text{CM}(t)$ (blue line) and of the low-pass filtered centre of mass (red line), both for $a_0=4$[mm], $f_0=55.7$[Hz], $S_d=200$ and $F=12$. (b) The autocorrelation function of the low-pass filtered centre of mass signal is plotted versus time for different number of layers and shaking strength $S_d=200$. The dashed line represents the fit of the maxima for $F=8$.}
  \label{fig:minmax}
\end{figure}

Now that we qualitatively understand the physics behind the LFO, we turn our attention to the driving mechanism. When one identifies an oscillation it is namely not sufficient to identify the oscillating object itself, but also what keeps the oscillation going, e.g., a bow needs to be drawn across a violin string in order for the latter to produce sound. To answer this question, we note that the LFO is quasi-periodic rather than purely harmonic, i.e., it is not a clear periodic signal in time. Moreover it appears to be decoupled from the driving in the sense that for constant $a_0f_0$ the amplitude and frequency of the LFO are independent of $a_0$ c.q. $f_0$ which can therefore be ruled out as the direct driving mechanism. We compute the autocorrelation function of the centre of mass 
\begin{equation}
\xi(t_0) = \frac{\langle z_\text{CM}(t+t_0)z_\text{CM}(t)\rangle_t}{\langle(z_\text{CM}(t))^2\rangle_t}\,,
\end{equation}
\noindent where $\langle...\rangle_t$ indicates a time average. The result is shown in Figure~\ref{fig:minmax} (a). Clearly, the autocorrelation function of the pure signal is dominated by the frequency of energy injection, which is the highest frequency present in the blue curve. Thus, the damped low-frequency oscillation that appears as a modulation of the high frequency signal is quite hard to recognise. We therefore choose to first low-pass filter the $z_\text{CM}(t)$ signal, with a cutoff frequency $f_c = f_\text{LFO} + (f_0-f_\text{LFO})/2$, that is, in between the frequencies of the LFO and the shaker.  The autocorrelation function of the filtered signal accurately captures the autocorrelation on a larger time scale, as shown in figure~\ref{fig:minmax} (a). The location of the maxima and minima of $\xi(t_0)$ are verified to correspond with the period $\tau_\text{LFO}=1/f_\text{LFO}$ of the LFO. Furthemore, we observe that the maxima (and minima) of $\xi(t_0)$ decay exponentially in time. The corresponding typical decay time scale $\tau$ is obtained by fitting the successive maxima to an exponential  $y(t_0) = e^{-t_0/\tau}$, as shown in figure~\ref{fig:minmax} (b). This time decay can be readily interpreted as a decorrelation time, i.e., it is a measure for how fast the oscillating system looses phase information, or, conversely, what the typical coherence time of the signal is. 

\begin{figure}[h!]
  \begin{center}
   \includegraphics[scale=1]{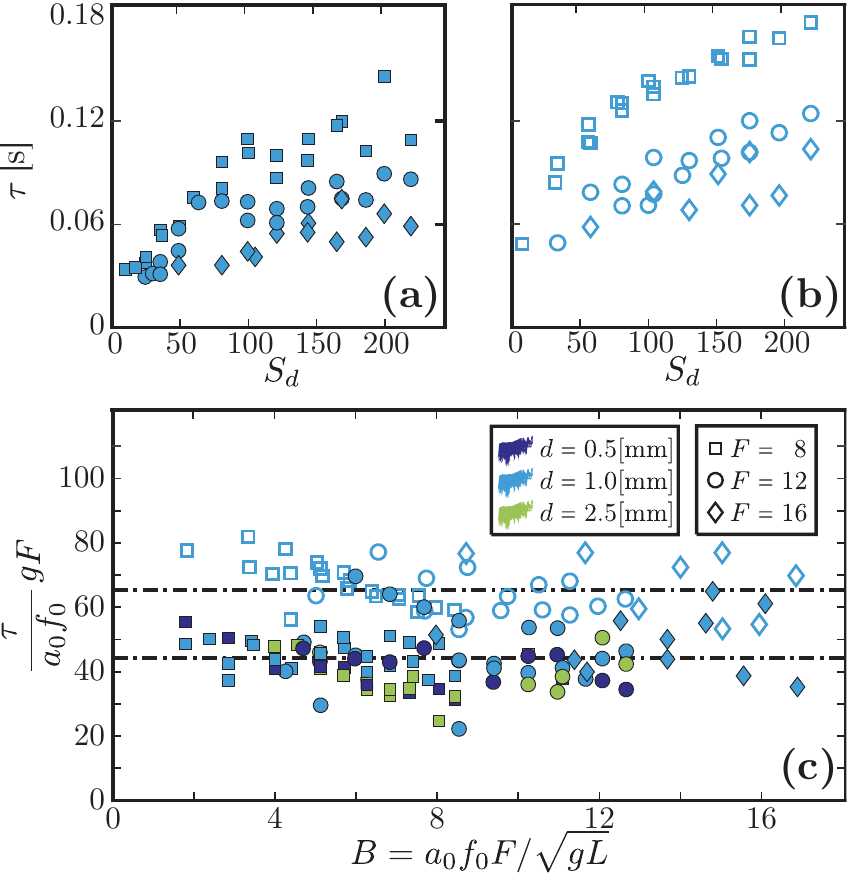}
  \end{center}
  \vspace{-0.2cm}
  \caption{Decorrelation time $\tau$ determined from the autocorrelation function $\xi(t_0)$ of the center of mass signal $z_\text{CM}(t)$. (a) Experimental data for $\tau$ as a function of the dimensionless shaking strength $S_d$ for different values of the number of layers $F$ and the particle diameter $d$ (see legend at the bottom). (b) Numerical data for $\tau$, again as function of $S_d$. (c) $\tau$ non-dimensionalized with the dissipative time scale $\tau_\text{d} = a_0f_0/(gF)$ plotted versus the shaking parameter $B$. The horizontal lines correspond to the averaged value of $\tau/\tau_\text{d}$ for the numerical simulations and the experiments. As in previous plots, different symbols correspond to different numbers of layers $F$ and different colors to different particle diameters $d$, and the solid and open symbols represent experimental and numerical data, respectively.}
\label{fig:tauvsSHL} 
\end{figure}

The decorrelation time increases with $S_d$, as was suggested by the narrowing peaks of the spectra of $z_\text{CM}(t)$ (figure \ref{fig:PSD}). Moreover, we see that the correlation is inversely proportional to both d and F, suggesting that, in general, correlations are higher for systems with a higher number of particles, as seen in figure \ref{fig:tauvsSHL} for experimental and numerical data.

 \subsection{Time scales} 
 
 In the literature three different time scales have been identified in (dilute) vertically driven granular systems from hydrodynamic descriptions~\cite{bromberg_development_2003,wakou_fluctuation-dissipation_2012}. These three time scales are associated to the different processes that take place in a driven granular system. The first and fastest inertial time scale is associated to the mechanical response of the system and is identical to the one we introduced to non-dimensionalize the LFO frequency, namely $\tau_\text{osc} = a_0f_0/g$. The second time scale is connected to heat conduction and was shown to be equal to $\tau_\text{heat}=F a_0f_0/g$ and, finally, the third and largest time scale corresponds to dissipative processes and can be approximated as $\tau_\text{diss} \sim  a_0f_0/(g \varepsilon F)$. Here, the inelasticity coefficient $\varepsilon = 1-e^2$ is given in terms of the (nominal) normal coefficient of restitution $e$. Since, although unknown, $\varepsilon$ can be assumed to be constant in the experiments conducted in this work we just use $\tau_\text{d} \equiv\sum\tau_\text{diss}=  a_0f_0/(g F)$ as a reference time scale.

When the decorrelation time  {$\tau$} is rescaled with $\tau_\text{d}$ it is roughly independent of any parameter, as can be appreciated in figure~\ref{fig:tauvsSHL} (c), where the dimensionless decorrelation time $\hat\tau\equiv\tau/\tau_\text{d}$ is plotted versus the shaking parameter $B$. Therefore,
\begin{equation}
\tau \sim \tau_\text{d} \sim \frac{a_0f_0}{gF}\,,
\end{equation}
\noindent which implies that the system becomes more coherent with increasing shaking strength (i.e., when $a_0f_0$ increases) and looses its coherency when the number of layers in the system $F$ increases. This is consistent  with the view that the LFO is driven by the noise in the system, which increases with the driving strength $a_0f_0$ but decreases when the number of layers $F$ in the system increases.   

Note that the constant value in figure~\ref{fig:tauvsSHL} is larger for numerical simulation than experimental data, which can be attributed to differences between simulation and experiment as, e.g., the role of particle-wall dissipation. In addition, its large value ($\sim50$) is connected to the fact that $\tau_d$ needs to be divided by the inelasticity $\epsilon$, which is typically much smaller than one, to obtain the dissipative time scale $\tau_\text{diss}$.

\subsection{LFO intensity}  
 
Finally, we turn to the strength of the oscillation by measuring the oscillation intensity $I_\textrm{LFO}$, namely the amplitude of the PSD at the oscillation frequency $f_\textrm{LFO}$.  This quantity is plotted in Fig. 10 for (a) the experiments  and (b) the numerical simulations as a function of the shaking strength $S_d$. We observe that $I_\textrm{LFO}$ (for any fixed $F$ and $d$) increases with the shaking strength $S_d$, and that there is a dramatic influence of the number of layers $F$, with $I_\textrm{LFO}$ dropping quickly as a function of $F$.

The shape of the PSD (Fig.~5) and the time-autocorrelation function (Fig. 8) suggest an approximate description in terms of a Langevin equation of a noise-driven harmonic oscillator (also known as Brownian motion in a harmonic potential) which is textbook material in nonequilibrium statistical mechanics \cite{zwanzig_2001}. 
\begin{equation}
\ddot{\zeta} = - \eta \dot{\zeta} - \omega_\textrm{LFO}^2\zeta + \xi\,,
\label{eq:Langevin}
\end{equation}
where $\omega_\textrm{LFO} = 2\pi f_\textrm{LFO}$, $\zeta(t) = z_\textrm{CM}(t) - \langle z_\textrm{CM} \rangle$ is the vertical deviation of the center of mass location from its time average $\langle z_\textrm{CM} \rangle$, $\eta = 2/\tau$ corresponds to twice the inverse decorrelation time, and $\xi(t)$ is delta-correlated white noise, obeying a fluctuation-dissipation relation $\langle \xi(t)\xi(t') \rangle = 2\eta V_T^2 \delta(t-t')$. Here, $V_T$ stands for the thermal velocity of the center of mass. In fact, such a Langevin approach had been suggested by Wakou and Isobe in Ref. [30], where a Langevin description of a dilute vibrated granular gas (i.e., without density inversion and the clearly observable Leidenfrost oscillation) was proposed. 

The PSD $S(\omega)$ corresponding to Langevin Eq.~(\ref{eq:Langevin}) is readily calculated theoretically~\cite{zwanzig_2001}
\begin{equation}
S(\omega)  = \frac{2\eta V_T^2}{[(\omega_\textrm{LFO}^2 - \omega^2)^2\,\,+\,\,\eta^2\omega^2]}\,.
\label{eq:PSD}
\end{equation}

\noindent Using the measured values for $I_\text{LFO}$, $f_\text{LFO}$ and $\tau$ we plot the above expression together with the experimental PSDs in Fig. 5 and find very good agreement for frequencies around and lower than the Leidenfrost oscillation frequency, but also that Eq. \ref{eq:PSD} significantly underpredicts the participance of the higher frequencies in the experimental PSD.

Clearly, the numerator being constant, $S(\omega)$ is maximum when the denominator obtains its smallest value, which gives:
\begin{equation}
I_\textrm{LFO}  = \frac{2V_T^2}{\eta(\omega_\textrm{LFO}^2 - \tfrac{1}{4}\eta^2)} \approx \frac{\tau V_T^2}{\omega_\textrm{LFO}^2}\,,
\label{eq:ILFO}
\end{equation}
where the last approximation originates from the expectation that $\eta \ll \omega_\textrm{LFO}$. The only quantity of which the scaling behavior is unknown is the thermal velocity of the center of mass, $V_T$.  

Under the rather bold assumption of energy equipartion, $V_T$ is linked to the thermal velocity $v_T \sim a_0f_0$ of the grains by $\tfrac{1}{2}M_{tot}V_T^2 \approx  \tfrac{1}{2} m v_T^2$, with the total mass $M_{tot} = Nm$ given by the product of the number of particles $N$ and the grain mass $m$. This leads to
\begin{equation}
V_T^2 \approx \frac{m}{M_{tot}}v_T^2 \sim \frac{a_0^2f_0^2}{N}\,.
\label{eq:V_T}
\end{equation}
Using the above result with the previously corroborated $\tau \sim a_0f_0/(gF)$ and $\omega_\textrm{LFO} \sim g/(a_0f_0)$ provides the scaling   
\begin{equation}
I_\textrm{LFO}  \sim  \frac{(a_0f_0)^5}{g^3 F N} \sim  \frac{(a_0f_0)^5 d^2}{g^3 F^2 L^2} 
\label{eq:ILFOscaling}
\end{equation}
where in the last step we have used that $N=FL^2/d^2$. In Fig. 10 (c) we rescale $I_\textrm{LFO}$ with the right hand side of Eq.~(\ref{eq:ILFOscaling}) and plot it as a function of the shaking parameter $B$ and find a fair data collapse for both the experimental and the numerical data. Moreover it can be seen that the dimensionless $I_\textrm{LFO}$ data decays exponentially with $B$. Since $\hat{\tau}$ is approximately constant and $f_\textrm{LFO}$ has non-exponential behavior (Fig. 7 suggests that it increases as a power law of $B$), this exponential decay appears to be connected to the violation of energy equipartition, as might be anticipated for this far-from-equilibrium system. More specifically, we find that the ratio of the thermal energy of the center of mass and that of the particles decays exponentially, i.e.,
\begin{equation}
V_T^2 = 4\pi^2\frac{a_0^2f_0^2}{N} \phi(B) \,,
\label{eq:V_T2}
\end{equation}
where $\phi(B)$ is an exponentially decaying function of the shaking parameter $B$.

\begin{figure}[h!]
  \begin{center}
   \includegraphics[scale=1]{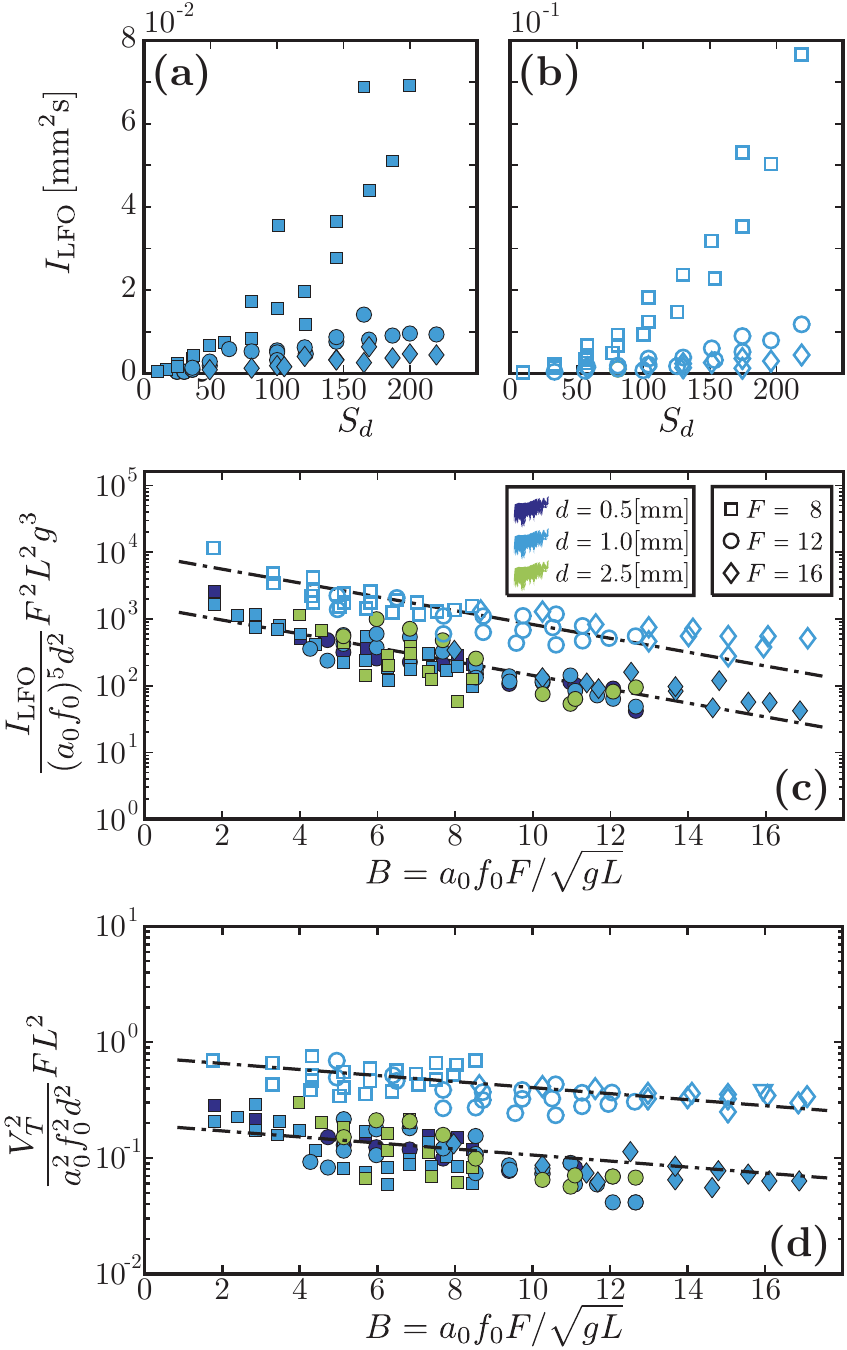}
  \end{center}
  \vspace{-0.2cm}
  \caption{Intensity $I_\text{LFO}$ obtained from the PSD of the center of mass signal $z_\text{CM}(t)$. Top left: Experimental data for $I_\text{LFO}$ as a function of the dimensionless shaking strength $S_d$ for different values of the number of layers $F$ and the particle diameter $d$ (see legend at the bottom). Top right: Numerical data for $I_\text{LFO}$, again as function of $S_d$. Bottom: $I_\text{LFO}$ non-dimensionalized by $(a_0f_0)^5d^2/(L^2g^3F)$ plotted versus the shaking parameter $B$. The dashed-dotted line serves as a guide to the eye. As in previous plots, different symbols correspond to different numbers of layers $F$, different colors to different particle diameters $d$, and the solid and open symbols represent experimental and numerical data, respectively.}
\label{fig:Icm} 
\end{figure} 

To test the above, we use Eq. (12) to express $V_T^2$ in terms of the measured quantities $I_\text{LFO}$, $f_\text{LFO}$ and $\tau$, subsequently non-dimensionalize with $a^2_0f^2_0/N$ and obtain the result plotted in Fig. 10 (d), which corroborates our expectation: there indeed appears to be an exponential decay of the ratio of the thermal energy of the center of mass and that of the particles. There are many probable causes why energy equipartition may deteriorate for larger $B$. The most obvious one is dissipation, which makes it unlikely for the thermal energy of the gas particles and that of the plug to be equilibrated. A more subtle effect is that the plug is not a fixed entity moving as a Browian particle, but itself consists of many particles that will respond to collisions with the particles in the gaseous layer. This will lead to the observed much stronger participation of the higher frequencies in the PSD than expected based on the simple Langevin model Eq. (10). And a stronger participation of the higher frequencies directly implies a lower intensity at the resonance frequency $f_\text{LFO}$, even if there would be
equipartition.


\section{Conclusions}\label{s:Conclusions}

We have experimentally and numerically studied the low-frequency oscillation (LFO) in a vibrated quasi-one-dimensional column of granular material in the Leidenfrost state. This LFO manifests itself as a vertical oscillation of the dense top layer of the Leidenfrost state. We determined the LFO frequency from the power spectral density of the centre of mass signal, for different amplitude and frequency of the driving, particle diameters and number of layers in the system.

In search of scaling laws, we constructed a simplified mass-spring model of the Leidenfrost state and argued that the LFO frequency should be inversely proportional to the fast inertial time scale, which is independent of the particle diameter and the number of layers. The dimensionless oscillation frequency is found to depend on the shaking parameter $B = a_0f_0F/\sqrt{gL}$, which was motivated qualitatively from the balance of energy input and dissipation in the dense layer. Experimental and numerical results are in fair agreement with each other.

Subsequently, we studied the time autocorrelation function of the centre of mass signal, and found that the LFO decorrelates with a typical decay time $\tau$, which was found to be proportional to the dissipative time scale $\tau_\text{diss}=a_0f_0/(gF)$, and otherwise independent of any other parameter in the system. Such a decorrelation is consistent with the view that the LFO is driven by the inherent noise in the system, which increases with the driving strength $a_0f_0$ but decreases with the number of layers $F$.
 
 Finally, we argue that the vertical position of the center of mass can be approximately described by a Langevin equation with weak white noise. We find that the amplitude of the oscillation, measured as the oscillation intensity in the PSD, scales consistently with that expected from the Langevin description. We find that energy equipartition is violated in this far-from-equilibrium system and that the portion of energy partitioned with the oscillating dense top layer is decaying significantly with increasing shaking strength.


\bibliographystyle{ieeetr}
\bibliography{GranularLeidenfrostEffect}{}

\end{document}